# Partial Hydrogenation of N-heteropentacene: Impact on molecular packing and electronic structure

Yutaro Ono,[1] Ryohei Tsuruta,[1] Tomohiro Nobeyama,[1] Kazuki Matsui,[1] Masahiro Sasaki,[1] Makoto Tadokoro,[2,4] Yasuo Nakayama,[3,4] and Yoichi Yamada[1]

*1. Faculty of Pure and Applied Sciences, University of Tsukuba, 1-1-1 Tennodai, Tsukuba, 305-8573 Japan*

*2. Department of Chemistry, Faculty of Science, Tokyo University of Science, Kagurazaka 1-3, Shinjuku-ku, Tokyo 162-8601, Japan*

*3. Department of Pure and Applied Chemistry, Tokyo University of Science, 2641 Yamazaki, Noda 278-8510, Japan*

*4. Division of Colloid and Interface Science, Tokyo University of Science, 2641 Yamazaki, Noda 278-8510, Japan*

**ABSTRACT**

Four-nitrogen-containing 5,6,13,14-Tetraazapentacene (BTANC) has attracted attention as a new n-type organic semiconductor with a rigid crystalline phase due to intermolecular C−H···N hydrogen bonding. However, in the thin film transistor of BTANC, poor carrier transport properties and low stability in the ambient condition have been reported so far; thus further refining and understanding of the thin film of BTANC will be required. Here, by means of carefully-controlled vacuum deposition of BTANC in the narrow window of temperature avoiding impurity sublimation and thermal degradation of molecules, we produced a well-defined monolayer on Cu(111) for molecular-level investigations. Synchrotron photoemission of the monolayer revealed a noticeable alteration of the chemical state of N atoms, which is unexpected for the pure BTANC molecule. In addition, molecular imaging of the monolayer by scanning tunneling microscope (STM) revealed that the molecular packing structure in the monolayer significantly differed from that in the single crystal of BTANC. These observations can be interpreted as a result of the partial hydrogenation of N atoms in BTANC and the emergence of the N−H···N type intermolecular hydrogen bonding in the monolayer. These findings will provide a general remark and strategy to control the molecular packing structure and electronic property in the molecular films of the nitrogen-containing acenes, by means of controlled hydrogenation.

**INTRODUCTION**

N-heteroacenes have attracted attention in terms of manipulating the electronic state of acenes to n- (or electron acceptor) type, [1–6] as well as stabilizing the assembly structure due to the intermolecular hydrogen bondings between hydrogens and nitrogens.[1,7–12] Theoretical investigations have suggested that the electron affinity of N-heteropentacenes can be significantly altered by the substitution of N atoms and showed that the conduction polarity alters into n-type.[7,13,14] The molecular packing properties of these molecules can also be tuned by changing the number, position, and valence state of N atoms in the pentacene backbone.[7,12,15,16] Both the experimental and the theoretical works have demonstrated that N-heteropentacenes with N−H···N or C−H···N hydrogen bondings show dense molecular packing.[1,7–12] Miao *et al.* utilized these properties of N-heteropentacene to fabricate an organic thin film transistor (OTFT) with hole mobility 0.45 $cm^2/Vs$,[17] which is comparable with that of amorphous silicon TFTs. Analysis of crystal structure by X-ray diffraction revealed that the molecular packing structure of their thin film differed from that of pentacene thin films in previous studies.[17–19]

Among N-heteroacenes, in the present study, we have focused on 5,6,13,14-Tetraazapentacene (BTANC, Figure 1a). BTANC has four nitrogen atoms, and was synthesized relatively recently by Isoda *et al.*[8] They clarified that the single crystalline structure of BTANC indeed showed dense molecular packing due to C−H···N hydrogen bondings, with BTANC molecules packed with antiparallel fashion, as shown in Figure 1a.[8] They also fabricated the OFET using BTANC and confirmed the electron transport properties, although the electron mobility was as small as ~$10^{-4}$ $cm^2/Vs$. They also noted that the OFET was degraded within 30 minutes in the ambient condition.[8] The small mobility and low air stability of BTANC is contradicting to the expected features of the N-heteropentacene. However, since BTANC is a relatively new molecule, it is still possible that the film of BTANC has not yet been optimized in terms of purity and the film structure, to which the mobility and stability are quite sensitive. In addition, N-heteropentacenes always have a possibility of hydrogenation of nitrogen atom(s), and this can also alter the electronic and structural properties of the solid films.[17,20] However, there have been only few experimental works on the film state of BTANC[5,8], and the details of the structure and electronic properties of the thin film of BTANC have remained elusive.

In the present work, we aimed to fabricate pure and well-ordered film of BTANC for detailed and molecular-level characterization of the solid film. For this purpose, we demonstrated a controlled deposition of BTANC thin film preventing a contamination of impurities and a thermal denaturation of BTANC during deposition. The core-level photoemission spectra of the fabricated monolayer on Cu(111) revealed the significant splitting of the N-1s photoemission peak, indicating that the nitrogen atoms of BTANC are partially hydrogenated, and this conclusion was also supported by the DFT calculations. At the same time, in the molecular-resolved image of the monolayer, it was seen that the molecular arrangement of BTANC is significantly different from that in the single crystalline; most of the molecules are arranged in the "parallel" way, and quite a few molecules in the monolayer took the anti-parallel ordering. The partial hydrogenation of BTANC in the monolayer can straightforwardly explain the parallel ordering of the molecules by means of the newly emerged N−H···N hydrogen bonding. Thus, it is shown that the partial hydrogenation of BTANC alters not only the electronic structure of a single molecule, but also the molecular packing structure of the film. These findings are of general importance for tailoring the suitable molecular film of the N-heteroacenes for application in organic electronics, by controlling the hydrogenation.

**EXPERIMENTAL METHOD**

BTANC was synthesized using a procedure detailed in a previous study[8]. Electronic states of BTANC dissolved in $CHCl_3$ were evaluated via UV-visible spectrophotometry (UV-Vis) with JASCO V-630 UV-Visible Spectrophotometer. Preparation and investigation of structure of BTANC monolayer are conducted using an in-situ STM system.[21,22] The substrates for molecular layers were single crystalline Cu(111) cleaned by repeated Ar ion sputtering and annealing cycles. The cleanliness of the substrate was confirmed by low-energy electron diffraction (LEED) and direct observation using STM. Molecular layers of BTANC were fabricated through a vacuum deposition with a homemade Tantalum Knudsen cell with a thermocouple, which enabled accurate cell temperature control and stable deposition. Sublimation rate during deposition were monitored using a quartz crystal microbalance (QCM). Structures of molecular layer were investigated using LEED and STM. Electronic states of molecular layer were examined via in-situ PES performed at beamline BL13B of KEK-PF facility with an excitation photon energy of 1500 eV. All measurements of molecule layer were performed at room temperature in ultra-high vacuum (~$10^{-8}$ Pa). STM images were processed using the WSxM software.[23]

All quantum calculations were performed using Gaussian 16.[24] All geometries were optimized using density functional theory (DFT) at the m062x/6-311++G(d,p) level of theory without any symmetry assumptions for good calculation to reflect weak interactions.[25] Harmonic vibration frequency (Freq) calculation was performed at the same level of theory to verify all stationary points as local minima (with no imaginary frequency). Zero-point energy, enthalpy, and Gibbs free energy correlation at 298.15 K and 1 atm were estimated with the results of the Freq calculations. Single point calculations were performed to estimate the energy of molecules, orbital energy, and optical property at the m062x/6-311++G(2d,p), CAM-B3LYP/6-311++G(2d,p), and B3LYP-D3/6-311++G(2d,p) Level of theory. B3LYP-D3/6-311++G(2d,p), showing the closet HOMO-LUMO gap compared to our experimental optical gap, was adopted to analyses of interaction energies and orbital energies. Optimized structures were visualized using GaussView 6. Cartesian coordinates of optimized structures were summarized in this supplementary material.

**RESULTS AND DISCUSSION**

Firstly, we confirmed the absorption spectra of the powder of BTANC dissolved in the chloroform solution (Figure 1b). The absorption spectrum was fully consistent with the previous studies.[5,8] The maximum absorption peak was observed around 450 nm, and Isoda *et al*. attributed the onset of this peak to the optical gap of BTANC of 2.68 eV.[8] In addition to this peak, there is a weak intensity at around 600 nm, which has been also observed in previous studies. Isoda et al. attributed this peak to the CT band,[8] while Hoffmann et al. attributed this peak to the optical gap (1.6 eV) and showed good correspondence with the EELS results.[5]

Then, we optimized the condition of the vacuum deposition of BTANC. Figure 1c shows the measured thickness at the QCM, when the BTANC cell was heated with a nearly constant increase rate of temperature (approximately 2.5 °C/min). The sublimation rate started increasing from around 100 °C and rose rapidly to about 200 °C. The rate became quite small for the cell temperatures above 350 °C. It is considered that BTANC is sublimated in the temperature range of 250-300°C, while the molecule can degrade at elevated temperatures. We also recognized that the color of the BTANC powder in the cell changed into black after heating to 400 °C and the black powder was not solvable to chloroform solution, suggesting that the BTANC can be denatured above 350 °C.

Figure 2a shows the X-ray photoelectron spectra of a monolayer prepared with two different temperature conditions of the cell to investigate the presence of impurities; the green line in Figure 2a is the XPS spectrum of monolayer deposited at the lower temperature of 150 °C, which is slightly lower than the BTANC sublimation temperature, and the blue line is monolayer deposited at 250 °C after degassing the possible impurities by preheating at the cell temperature of 150 °C for about 3 hours. The two spectra are not very different, but the green spectrum shows the appearance of the characteristic O 1s peak, as shown in the inset. Since the pure BTANC molecule

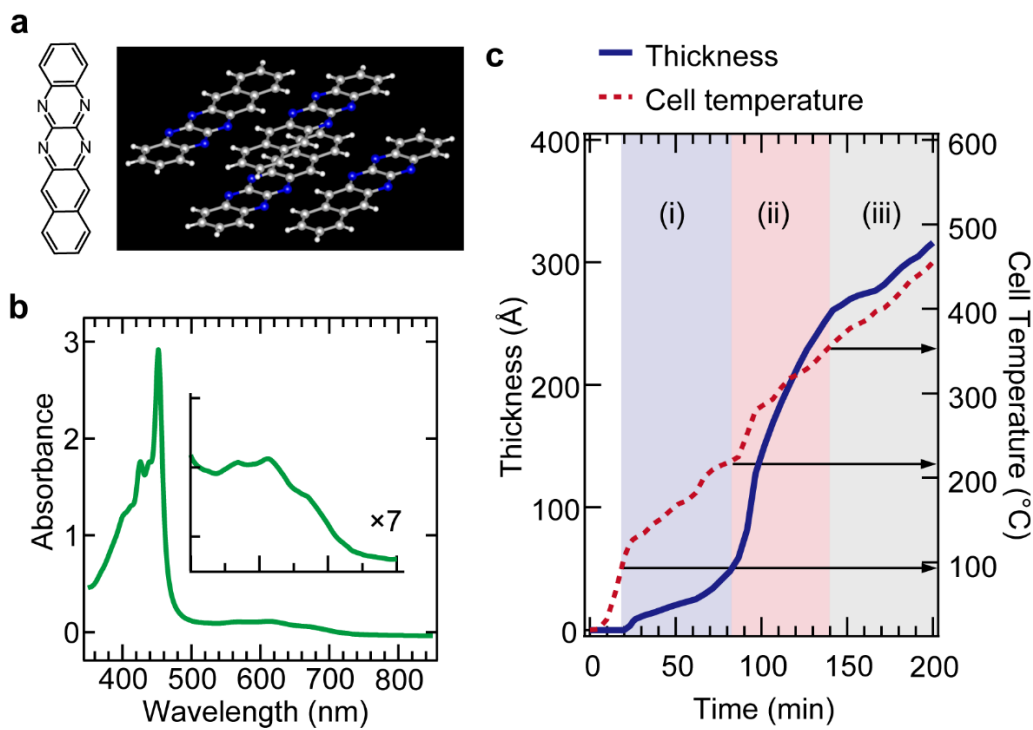

**Figure 1** (a) The structural fomula (left) and single crystal structure of BTANC (right). (b) Absorption spectra of BTANC in CHCl$_3$. (c) The increase of thickness of deposited film and cell temperature as function of deposition time. Upper photographs are solid states before heated (As-Is), after heated in (ii) 250°C and (iii) 400°C for approximately one hour.

does not contain oxygen, the presence of oxygen should be due to the contamination of oxygen and water in the air. Therefore, the increase in the thickness observed around 100 °C in figure 1b should be due to the sublimation of this oxygen-containing impurities. On the other hand, the oxygen peak disappeared in the blue (bottom) spectrum, suggesting that the impurity can be removed by the preheating. These facts suggest that the deposition of the pure BTANC is possible in the temperature window, above the sublimation of the oxygen-related impurity (150 °C) and below the degradation temperature of BTANC itself (350 °C).

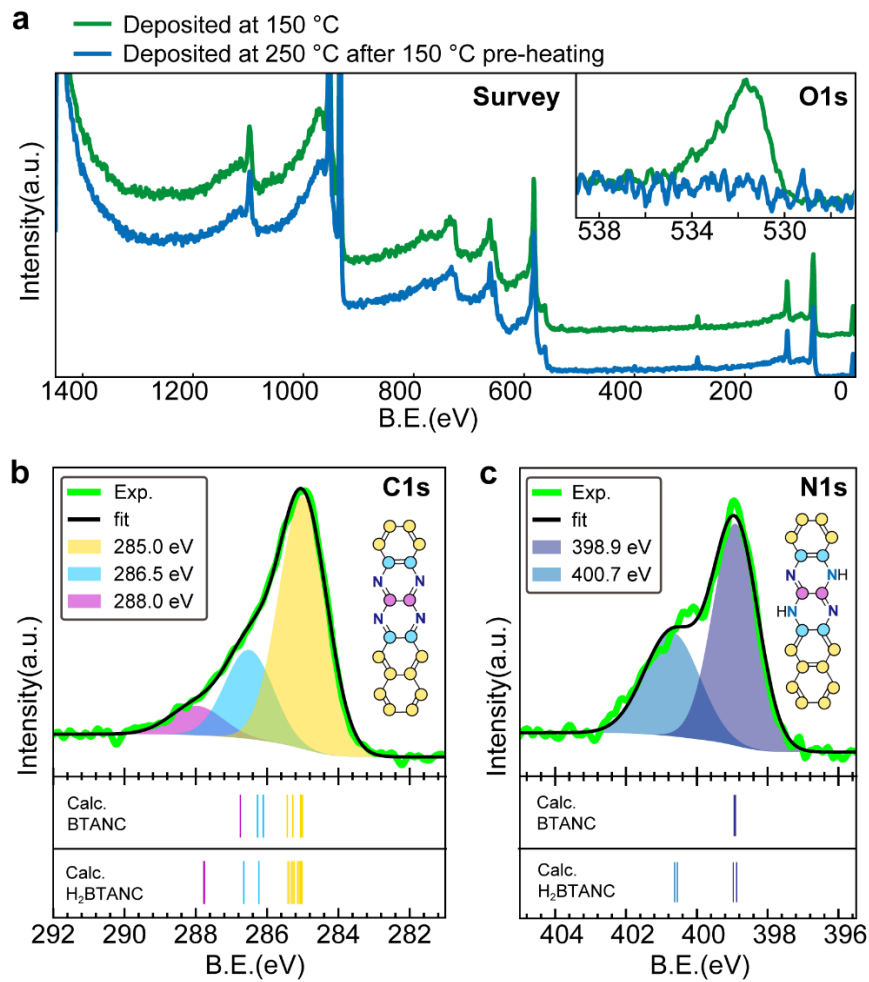

**Figure 2.** (a) Survey of XPS spectra of the films deposited in 150°C (green line) and 250°C after 3h preheating in 150 °C (dashed blue line). Inset is O1s spectra. (b, c) C 1s and N 1s XPS spectrum of BTANC on Cu(111). Backgrounds are estimated by Shirley background. Fitting components are marked by different colors. Quantum chemical calculation results of a single BTANC and H$_2$BTANC molecule as vertical bars, where the vertical scale is slid to align the energy position to the experimental peak position. The colors of bars BTANC and H$_2$BTANC correspond to the color coding of the carbon atoms in the molecular model in upper right of (b) and (c).

To analyze the electronic structure of the BTANC monolayer, detailed core-level photoemission spectra were obtained. Figure 2b shows the C 1s spectrum from the BTANC monolayer on Cu(111), prepared carefully in the temperature window of 250 ~ 300 °C after degassing impurities by preheating at 150 °C for 3 hours. The linewidth of the C 1s peak is found to be about twice that of the pentacene/Cu(111) in a previous study.[26,27] This broadening should reflect the presence of the C atoms with different chemical states in the molecule. BTANC molecule has three types of carbon atoms as shown in the model in Figure 2b: carbon atoms not adjacent to nitrogen atoms (yellow), carbon atoms adjacent to one nitrogen atom (blue), and carbon atoms being adjacent to two nitrogen atoms (purple). The C 1s photoemission peak can be indeed fitted by three Gaussian components as shown in Figure 2b. The FWHM of each component was set at 1 eV, according to the previous study of pentacene/Cu(111).[26,27] The main peak (yellow) is the lowest binding energy of 285.0 eV. This binding energy is close to that of C 1s of pentacene/Cu(111), therefore, this is considered to be due to the carbon atoms not being adjacent to the nitrogen atoms. Then, the subpeaks (blue and purple) at 286.5 eV and 288.0 eV can be attributed to the carbon atoms adjacent to the nitrogen atoms. The area intensity ratio of yellow: blue: purple peaks is 6.00:2.12:0.77, which is in good agreement with the stoichiometric ratio of yellow, blue and purple carbon atoms of 6:2:1 in the BTANC model. Therefore, blue and purple components are attributable to the blue and purple carbon atoms, respectively. The lower part of

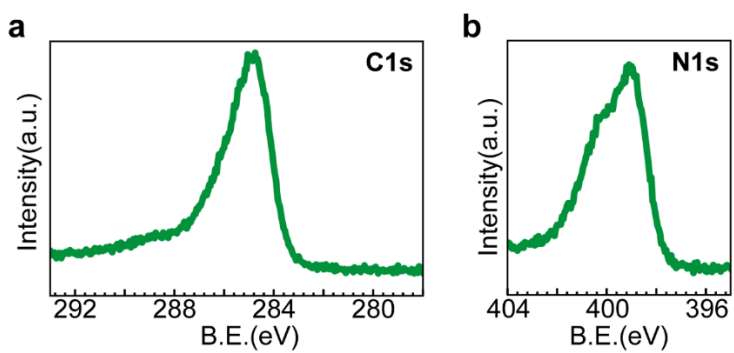

**Figure 3.** XPS spectrum of (a) C1s and (b) N1s from the powder sample on copper substrate. Each spectrum was normalized at each peak top.

Figure 2b shows the energy levels of the C 1s of BTANC determined from DFT calculations of the single molecule. The color of the vertical bar indicating the energy levels corresponds to the color coding of the carbon atoms. Note that, since DFT does not provide the absolute value of the binding energy, we put the calculated result so that the binding energy for the main peak coincides with the experimental peak position. It is seen that the binding energy of C 1s is blueshifted with the increasing number of adjacent nitrogen atoms due to electron transfer to nitrogen atoms. Thus, the result of the DFT calculations is qualitatively consistent with the C 1s spectrum obtained by photoemission spectroscopy. However, the experimental broadening was not quantitatively explained by DFT calculations of BTANC.

Figure 2c shows the N 1s spectrum of the monolayer, where the N 1s peak has a width of approximately 3 eV, which is also quite large. Fitting the N 1s spectrum required at least two Gaussian components whose energy splitting was approximately 1.5 eV. In a single molecule, due to the asymmetry of BTANC, the nitrogen atoms can be classified into two types, i.e., in the benzene side and the naphthalene side. However, the calculated difference in the binding energy of 1s for these N atoms is as small as 0.04 eV, as shown in the bottom panel of Figure 2c. Thus, the splitting of the N 1s peak in XPS should not be the property of BTANC itself. Similar splitting of N 1s has been reported in the XPS of the monolayer of 7,8,15,16-tetraazaterrylene (TAT), which contains four nitrogen atoms, deposited on Cu(111) and Ag(111) surfaces.[28] Even though the nitrogen atoms in TAT are in equivalent positions, the N 1s peak splits into two peaks by approximately 1.5 eV. Yang *et al.* attributed this splitting to the electron transfer to the one group of N atoms via the interaction with the substrate. In order to check the possibility of the substrate effect, we also examined photoemission from the powder sample (Figure 1c, As-is) put on the copper substrate. Figure 3 shows the C1s and N1s photoemission spectra from the powder sample. It is confirmed that the peak shape of C1s and N1s are similar to the case of the monolayer, and the N 1s peak of the powder sample also exhibited a clear split as in the case of the monolayer. This observation clearly suggests that the splitting of the N 1s in Figure 2c is not due to the substrate effect, but an intrinsic for the molecule. It is considered that the partial hydrogenation of BTANC is a possible reason for this chemical inhomogeneity of N atoms. Therefore, we calculated the partially hydrogenated BTANC using $H_2$BTANC as a model (Figure 2c), and the results of the simulation of C1s and N1s are shown in the bottom panels of Figure 2 b and c, respectively. The calculated C1s levels of $H_2$BTANC are considerably broader than that of BTANC and show better correspondence to the experimental C1s peak. Furthermore, the noticeable split of N1s is seen for H$_2$BTANC, also showing fair agreement with the experiment. It is seen that nitrogen atoms with hydrogen (blue) produces the peak at the high-binding-energy side of the nitrogen without hydrogen (purple), which is consistent with previous reports[29–31]. It is thus concluded that the BTANC is partially hydrogenated to H$_x$BTANC in the ambient condition, and the vacuum deposition of the powder also yields the film of H$_x$BTANC. It is noted that, according to computing study by Bunz *et al.*, the heat of hydrogenation of azapentacene has been shown to be generally negative. Especially in BTANC, the enthalpy change is as large as 41 kcal/mol, suggesting that hydrogenation can   proceed spontaneously.[3]

Figure 4a shows the STM image of the H$_x$BTANC monolayer on Cu(111) prepared with the same procedure as the samples used for XPS analysis. It was seen that the molecules formed a columnar structure in the monolayer. The molecular columns were elongated in three directions orthogonal to the [1-10] of the Cu(111) substrate. In the columns, molecules were packed approximately side-by-side manner and the direction of the molecular long axis was aligned in the [1-10] direction. The distance between molecules is 6.4 Å, which is close to the distance between molecules in single-crystal structures of BTANC (6.3 Å).[8] Figure 4c shows a 2-D Fourier transform image of Figure 4a. There exist regular spots in the center of the image, corresponding to the period of the molecular columns. The magnitude of the unit k-vector for these spots is approximately 0.5 nm$^{-1}$ corresponding to the separation of the columns of 1.8 nm. In addition to the spots, there exist line features forming hexagons. This feature corresponds to the periodicity of the adjacent molecules along the direction of columns, which was not fully periodic and therefore resulted in the line feature. The similar pattern is seen in the LEED image of the same sample as shown in Figure 4d.  This indicates that the molecular arrangement shown in Figure 4a spreads entire of the monolayer, and that the regular unit cell is absent in the entire monolayer.

The enlarged STM images show occupied state of the single molecules in the columns, obtained with negative sample bias voltage of -1.1 V.  The image shows an asymmetric shape in the long axis which is pinched at pyrazine rings. This image corresponds well to the HOMO-1 wave function of H$_2$BTANC or HOMO of BTANC (shown in Figure 4b). However, since the overall shape of the wavefunction of the frontier orbitals of H$_2$BTANC is similar to that of BTANC and also it does not strongly depend on the position of the added hydrogen atoms (Figure S2), we were not able to distinguish the molecular species from STM image. Nevertheless, we can determine the orientation of the molecule based on the STM image of the occupied state. In the enlarged STM image of a single molecular column, it is clearly seen that the molecules are mainly arranged in the same orientation, and we denote this arrangement as "parallel". It was also found that the adjacent molecules displaced each other in the perpendicular direction to the direction of the column, and the amount of this displacement was not uniform within the column. This irregular displacement of the molecules should be the origin of the line pattern in the Fourier-transformed image of the STM image and LEED pattern in Figures 4a and 4c. The parallel arrangement and random displacement of molecules in the monolayer are unexpected because the single crystalline BTANC consists of antiparallel and regular arrangement of molecules with strong hydrogen bonding. Therefore, the parallel arrangement and random displacement of molecules should be the characteristic property of the monolayer $H_x$BTANC.

In Figure 4e, the histogram of the distribution of apparent displacement between two adjacent molecules in the molecular column is shown. In the histogram of the displacement, we find two maxima at 0.0 and 2.5 Å, which are corresponding to the no-displacement configuration and the displacement of one benzene ring, respectively. It is seen that 24% of molecules are in the no-displacement configuration. This configuration is not stable for BTANC because of the lack of hydrogen bondings. However, it becomes possible for the partially hydrogenated $H_x$BTANC and one possibility is depicted in the bottom panel of Figure 4f, for $H_2$BTANC molecules (which we denote as Parallel $\Delta_0$). This pair allows two N−H···N hydrogen bondings without displacement of the molecule. On the other hand, more populations of the molecules are in the displaced configuration (Parallel $\Delta_1$). Although this configuration is also possible for BTANC, $H_2$BTANC molecules can form more stable pairs with larger numbers of hydrogen bondings, as shown in the example of the bottom panel of Figure 4f. In addition, the configuration of the pair with two benzene-rings displacement (Parallel $\Delta_2$) is also possible, as also shown in the bottom panel of Figure 4f. This pair exhibits a displacement of 4.7 Å, and the population of this pair was small. On the other hand, there existed small numbers of the molecular pair with two molecules in the opposite direction, which we referred to as an "Anti-Parallel" pair, as shown in Figure 4f. Although the non-hydrogenated BTANCs would be anti-parallel, the population of the pure BTANC in the monolayer is considerably small, considering that about 30-40% of all nitrogen is hydrogenated deduced from the XPS results. There can also be anti-parallel pair of the H₂BTANC with dihydrogenated on one side, as shown in Figure 4f. The calculated frontier orbitals of these molecules are curved and the appearance of the STM image of this pair fairly matches the calculation. This anti-parallel pair due to one-sided hydrogenated molecules appears as a dimer and tended to disrupt the elongation of the molecular column. However, the small population of this antiparallel pair suggests that the population of H$_2$BTANC with hydrogen atoms only on one side is small.

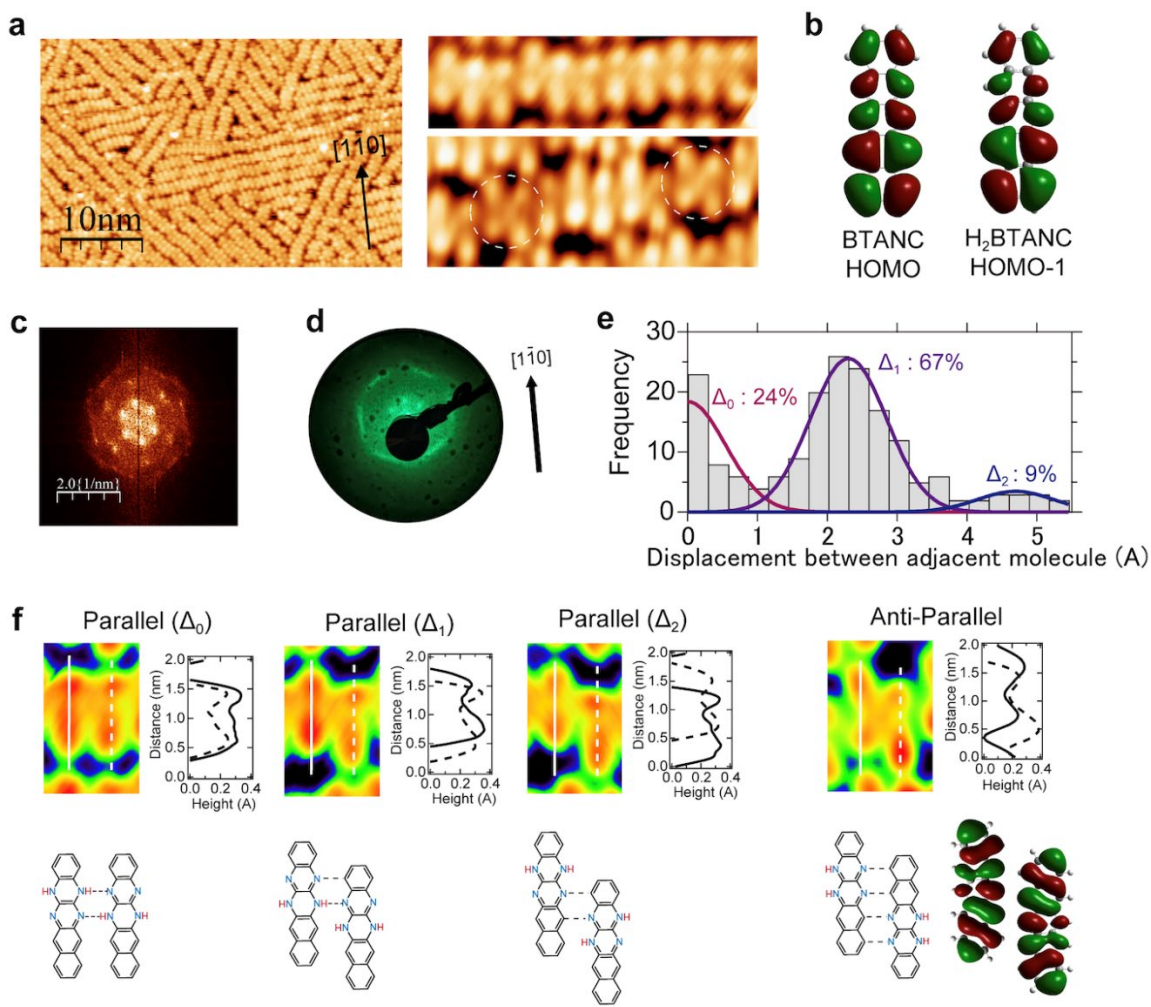

**Figure 4.** (a) STM image of monolayer of BTANC on Cu(111), and the [1-10] direction is indicated by a black arrow. Bias voltage = –1.1 V, setpoint = 0.4 nA. The right images are rows of molecules. The right-top image is 7.5 nm × 2.0 nm, and the right-bottom image is 7.5 nm × 3.0 nm. All molecules in the top image are same direction, while the bottom image includes two anti-parallel pair (marked by white circle). (b) HOMO of BTANC and HOMO-1 of H₂BTANC. (c) 2D Fourier transform of (a). (d) LEED pattern of monolayer with beam energy of 30 eV. (e) Histogram showing the distribution of apparent displacement between two adjacent BTANC molecules. (f) Classification of displacement between parallel and anti-parallel two adjacent molecules. STM images, and their line profiles and molecular models. Dashed lines in two molecular models mean hydrogen bonds. Height axis of each line profile are shifted since peaks fall between 0-4 Å.
12

Thus, the experimental results from both STM and photoemission can be fully explained by the partial hydrogenation of the BTANC; the partial hydrogenation of the BTANC resulted in both the chemical inhomogeneity of the nitrogen atoms and structural inhomogeneity in the film. Therefore, the control of the electronic structure and the molecular arrangement in the film should be possible by further controlling the hydrogenation of BTANC. Also, the results presented here for BTANC provides important aspect in the application of N-heteropantacenes in the advanced organic devices.

**CONCLUSIONS**

In this work, we fabricated the well-defined monolayer of BTANC via a controlled deposition and characterized it in detail. From the photoemission of the deposited film and material powder, we found that the BTANC molecules are partially hydrogenated to H$_x$BTANC. The characteristic molecular arrangement of the monolayer of H$_x$BTANC can be explained by considering the N−H···N hydrogen bonding between H$_2$BTANC molecules. Therefore, the present results suggested that both the electronic structure and the molecular arrangement in the film can be controlled if the partial hydrogenation H$_x$BTANC, such as position and the number of hydrogens, can be controlled. These issues of hydrogenation can be common aspects in the application of N-heteropentacenes in advanced organic devices.

**ACKNOWLEDGEMENT**

This work was supported by JSPS KAKENHI Grant Nos. 22K18268, 21H01805 and 20H02808. This work was partly performed under the approval of the Photon Factory Program Advisory

Committee (2021S2-003). A part of this study was supported by Tsugawa Foundation. Some of the calculations were performed via a cloud calculation system, Chempark (HPC Systems, Japan).

# Supporting Information

# Partial Hydrogenation of N-heteropentacene: Impact on molecular packing and electronic structure

*Yutaro Ono,[1] Ryohei Tsuruta,[1] Tomohiro Nobeyama,[1] Kazuki Matsui,[1] Masahiro Sasaki,[1] Makoto Tadokoro,[2,4] Yasuo Nakayama,[3,4] Yoichi Yamada[1]*

1. Faculty of Pure and Applied Sciences, University of Tsukuba, 1-1-1 Tennodai, Tsukuba, 305-8573 Japan

2. Department of Chemistry, Faculty of Science, Tokyo University of Science, Kagurazaka 1-3, Shinjuku-ku, Tokyo 162-8601, Japan

3. Department of Pure and Applied Chemistry, Tokyo University of Science, 2641 Yamazaki, Noda 278-8510, Japan

4. Division of Colloid and Interface Science, Tokyo University of Science, 2641 Yamazaki, Noda 278-8510, Japan

**Tabel S1.** Comparison table of calculated HOMO-LUMO gap ($E_g$) for BTANC monomar.

| Density functionals | Basis sets | $E_g$ (eV) |
|---|---|---|
| B3LYP-D3 | 6311++G(2d,p) | 2.23 |
| CAM-B3LYP | 6311++G(2d,p) | 4.36 |
| M062x | 6311++G(2d,p) | 3.99 |

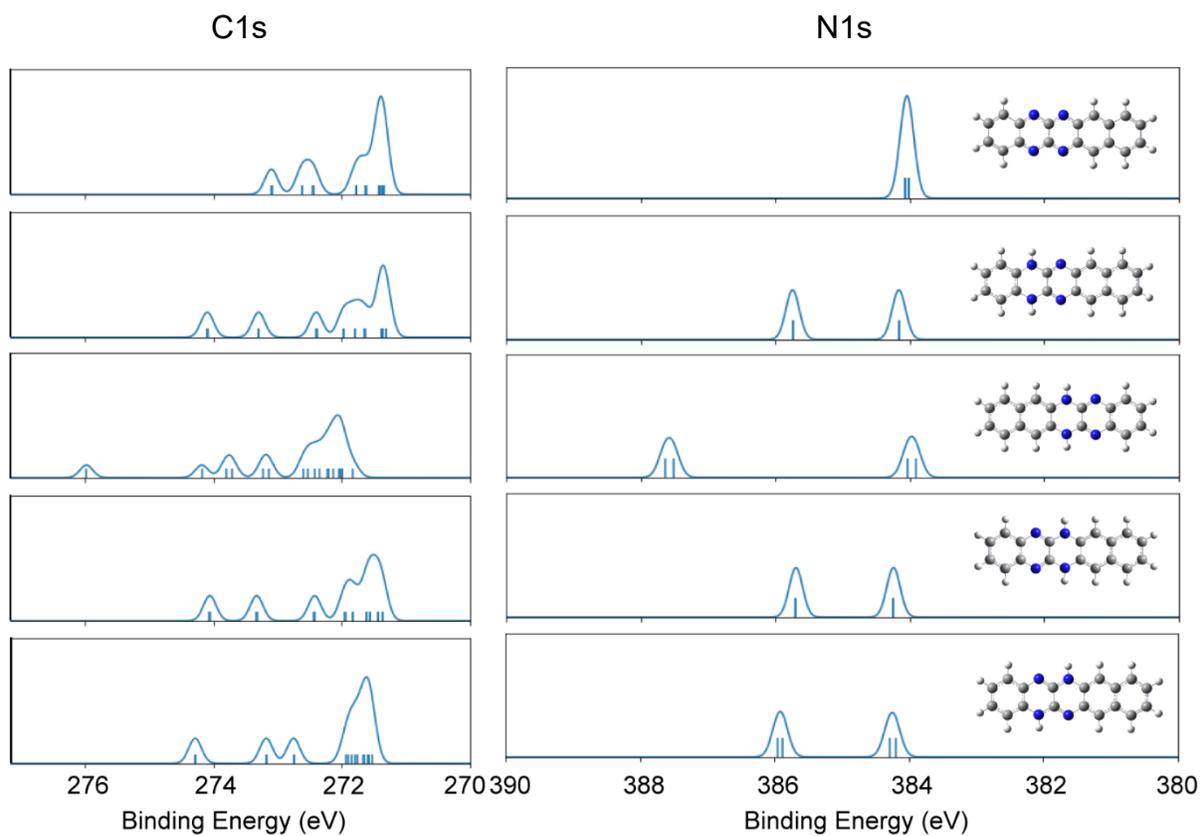

**Figure S1.** Quantum chemical calculation results of C1s and N1s of a single BTANC and H$_2$BTANC molecules.

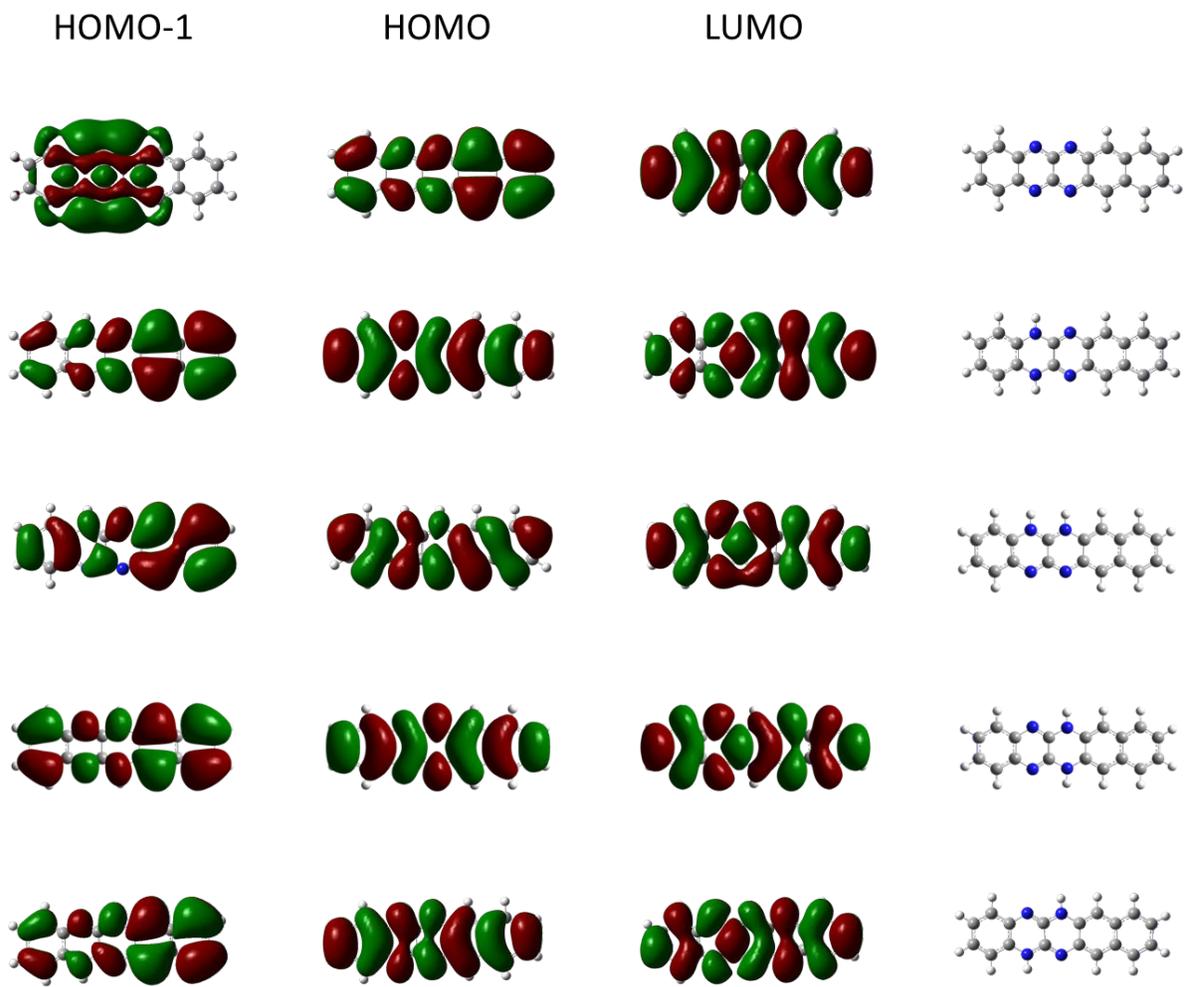

**Figure S2.** HOMO-1, HOMO, LUMO of BTANC and various H$_2$BTANCs.

**Appendix: Optimized cartesian coordinations obtained at the m062x/6-311++G(d,p) level of theory**

Optimized Cartesian Coordinates for Monomer of BTANC

| Atoms | Coordinates (Angstroms) | | |
|---|---|---|---|
| N | 0.00000000 | -0.05296100 | 1.42306700 |
| N | 0.00000000 | -2.34562200 | 1.42526500 |
| N | 0.00000000 | -2.34562200 | -1.42526500 |
| N | 0.00000000 | -0.05296100 | -1.42306700 |
| C | 0.00000000 | -1.18651900 | 0.72533800 |
| C | 0.00000000 | -3.45907400 | 0.72907700 |
| C | 0.00000000 | -4.72323400 | 1.42069700 |
| C | 0.00000000 | -5.88119800 | 0.72102800 |
| C | 0.00000000 | -5.88119800 | -0.72102800 |
| C | 0.00000000 | -4.72323400 | -1.42069700 |
| C | 0.00000000 | -3.45907400 | -0.72907700 |
| C | 0.00000000 | -1.18651900 | -0.72533800 |
| C | 0.00000000 | 1.08396500 | -0.72536700 |
| C | 0.00000000 | 2.31987900 | -1.41401300 |
| C | 0.00000000 | 3.51360600 | -0.72590000 |
| C | 0.00000000 | 4.78041900 | -1.41084300 |
| C | 0.00000000 | 5.94456700 | -0.71904700 |
| C | 0.00000000 | 5.94456700 | 0.71904700 |
| C | 0.00000000 | 4.78041900 | 1.41084300 |
| C | 0.00000000 | 3.51360600 | 0.72590000 |
| C | 0.00000000 | 2.31987900 | 1.41401300 |
| C | 0.00000000 | 1.08396500 | 0.72536700 |
| H | 0.00000000 | -4.69430300 | 2.50318800 |
| H | 0.00000000 | -6.83075200 | 1.24281800 |
| H | 0.00000000 | -6.83075200 | -1.24281800 |
| H | 0.00000000 | -4.69430300 | -2.50318800 |

| H | 0.00000000 | 2.29358200 | -2.49768000 |
|---|---|---|---|
| H | 0.00000000 | 4.77582300 | -2.49500400 |
| H | 0.00000000 | 6.89126500 | -1.24580100 |
| H | 0.00000000 | 6.89126500 | 1.24580100 |
| H | 0.00000000 | 4.77582300 | 2.49500400 |
| H | 0.00000000 | 2.29358200 | 2.49768000 |

Optimized Cartesian Coordinates for paralell model

| Atoms | Coordinates (Angstroms) | | |
|---|---|---|---|
| N | 2.81881900 | 0.38411000 | 0.13555700 |
| N | 4.77987300 | -0.79060000 | 0.29200500 |
| N | 6.24656100 | 1.64206700 | 0.05171000 |
| N | 4.27777000 | 2.80862300 | -0.12145100 |
| C | 4.14952300 | 0.39976500 | 0.15278000 |
| C | 6.09418400 | -0.76497300 | 0.31461500 |
| C | 6.82396100 | -1.99712200 | 0.46578000 |
| C | 8.17703200 | -1.98930800 | 0.49308700 |
| C | 8.91756900 | -0.75885400 | 0.37203400 |
| C | 8.28212100 | 0.42706400 | 0.22816000 |
| C | 6.84287600 | 0.47972600 | 0.19304800 |
| C | 4.89340100 | 1.63686700 | 0.02643300 |
| C | 2.94406500 | 2.79307200 | -0.14621200 |
| C | 2.23694800 | 4.00710300 | -0.31201400 |
| C | 0.85920200 | 4.02117600 | -0.35068000 |
| C | 0.11825900 | 5.24277900 | -0.53005300 |
| C | -1.23554700 | 5.23204800 | -0.57490700 |
| C | -1.96906400 | 4.00161500 | -0.44126000 |
| C | -1.32262700 | 2.82456100 | -0.26347200 |
| C | 0.11645400 | 2.78116700 | -0.21218200 |
| C | 0.78193900 | 1.58643600 | -0.04297100 |
| C | 2.19605800 | 1.55450300 | -0.01126900 |
| H | 6.24547300 | -2.90796000 | 0.55811100 |
| H | 8.72315500 | -2.91796500 | 0.60846700 |
| H | 9.99999600 | -0.79607000 | 0.39857100 |
| H | 8.81226000 | 1.36676300 | 0.13540400 |
| H | 2.81471500 | 4.91873900 | -0.41196900 |
| H | 0.67264900 | 6.16902300 | -0.63120700 |
| H | -1.78113700 | 6.15798500 | -0.71371400 |
| H | -3.05165600 | 4.02940100 | -0.48874300 |
| H | -1.85544400 | 1.88507600 | -0.15619500 |
| H | 0.24123900 | 0.65019400 | 0.06293400 |
| N | -4.15255300 | -2.85487000 | -0.24064300 |
| N | -2.15523700 | -3.97349100 | -0.40140000 |
| N | -0.75909900 | -1.53563300 | 0.06922900 |
| N | -2.75575100 | -0.42449800 | 0.23679300 |
| C | -2.82417800 | -2.81156800 | -0.20292800 |
| C | -0.84420500 | -3.91917300 | -0.36348100 |
| C | -0.08160600 | -5.12484500 | -0.56505900 |
| C | 1.27032800 | -5.08498200 | -0.52961400 |
| C | 1.97704000 | -3.84962900 | -0.29213900 |
| C | 1.31622400 | -2.68542100 | -0.09501800 |
| C | -0.12415800 | -2.66960700 | -0.12286100 |
| C | -2.11249600 | -1.57157800 | 0.03756300 |
| C | -4.09135100 | -0.46602200 | 0.20886200 |
| C | -4.82915800 | 0.71915200 | 0.42755500 |
| C | -6.20802400 | 0.71073700 | 0.40117000 |
| C | -6.97592600 | 1.90729700 | 0.62450900 |
| C | -8.32993800 | 1.88034000 | 0.59326900 |
| C | -9.03463700 | 0.65406000 | 0.33664500 |
| C | -8.35826200 | -0.49977400 | 0.12243600 |
| C | -6.91926800 | -0.52780600 | 0.14509400 |
| C | -6.21504800 | -1.69423200 | -0.06537700 |
| C | -4.80153700 | -1.70505000 | -0.03891800 |
| H | -0.63702100 | -6.03757300 | -0.74090000 |
| H | 1.84224900 | -5.99291000 | -0.68142500 |
| H | 3.06048500 | -3.85695900 | -0.27181400 |
| H | 1.83148800 | -1.74711700 | 0.08275300 |
| H | -4.27835200 | 1.63202200 | 0.62595100 |
| H | -6.44132000 | 2.83022700 | 0.81965300 |
| H | -8.89684400 | 2.78783300 | 0.76294600 |
| H | -10.11773500 | 0.66288800 | 0.31738900 |
| H | -8.88494300 | -1.42761300 | -0.07025000 |
| H | -6.72286900 | -2.63309400 | -0.25431200 |

Optimized Cartesian Coordinates for anti-parallel model

| Atoms | Coordinates (Angstroms) | | |
|---|---:|---:|---:|
| N | 2.03279400 | 0.85248900 | -0.36507200 |
| C | 3.36310000 | 0.88499100 | -0.31484500 |
| C | 1.38597500 | 1.98206500 | -0.06947400 |
| N | 4.02190300 | -0.25949500 | -0.61110700 |
| C | 5.33513700 | -0.22166200 | -0.56344800 |
| C | 4.08092300 | 2.09168100 | 0.04265600 |
| N | 5.43336800 | 2.11143000 | 0.08932600 |
| C | 6.05617900 | 0.99227900 | -0.20455000 |
| N | 3.44044500 | 3.22029300 | 0.34203700 |
| C | 2.10800500 | 3.19040000 | 0.29463800 |
| C | 6.08854200 | -1.40816300 | -0.87555600 |
| C | 7.44112800 | -1.38622700 | -0.83017100 |
| H | 5.52833300 | -2.29405100 | -1.14959400 |
| C | 8.15541200 | -0.18681500 | -0.47158300 |
| H | 8.00714700 | -2.27892300 | -1.06830400 |
| C | 7.49581700 | 0.95611600 | -0.17164700 |
| H | 9.23821000 | -0.21201200 | -0.44687500 |
| H | 8.00659200 | 1.87177100 | 0.09921000 |
| C | 1.37686900 | 4.35999100 | 0.60920500 |
| C | -0.00067500 | 4.36261900 | 0.57326300 |
| H | 1.93718500 | 5.24740400 | 0.88017300 |
| C | -0.76499700 | 5.53903300 | 0.89727500 |
| C | -0.71850600 | 3.15586000 | 0.20268300 |
| C | -2.11877100 | 5.51645900 | 0.85993200 |
| H | -0.22851900 | 6.43985500 | 1.17324400 |
| C | -2.82733200 | 4.31960500 | 0.49236300 |
| H | -2.68339700 | 6.40735900 | 1.10855600 |
| C | -2.15861500 | 3.18540100 | 0.17335400 |
| H | -3.91085400 | 4.33620900 | 0.47605000 |
| H | -2.68132800 | 2.27721100 | -0.11110900 |
| C | -0.02798400 | 2.00448700 | -0.10872700 |
| H | -0.56184500 | 1.09957800 | -0.38202400 |
| N | -3.44037700 | -3.22028600 | 0.34257500 |
| C | -4.08092400 | -2.09174500 | 0.04307600 |
| C | -2.10794100 | -3.19035200 | 0.29504400 |
| N | -5.43336300 | -2.11151600 | 0.08992700 |
| C | -6.05623800 | -0.99243500 | -0.20407800 |
| C | -3.36318100 | -0.88509900 | -0.31473400 |
| N | -4.02204800 | 0.25929900 | -0.61118200 |
| C | -5.33527500 | 0.22144800 | -0.56334200 |
| N | -2.03288200 | -0.85254700 | -0.36506100 |
| C | -1.38599200 | -1.98203900 | -0.06930400 |
| C | -7.49587200 | -0.95629600 | -0.17097500 |
| C | -8.15553700 | 0.18655200 | -0.47107200 |
| H | -8.00658700 | -1.87190400 | 0.10015800 |
| C | -7.44133300 | 1.38589900 | -0.83003600 |
| H | -9.23833200 | 0.21173000 | -0.44621300 |
| C | -6.08875300 | 1.40785700 | -0.87562000 |
| H | -8.00740700 | 2.27852700 | -1.06829200 |
| H | -5.52860700 | 2.29369700 | -1.14993600 |
| C | 0.02796400 | -2.00438900 | -0.10862600 |
| C | 0.71856200 | -3.15568400 | 0.20290300 |
| H | 0.56176900 | -1.09948600 | -0.38205700 |
| C | 2.15867100 | -3.18513100 | 0.17349300 |
| C | 0.00081700 | -4.36242900 | 0.57368600 |
| C | 2.82747300 | -4.31925200 | 0.49262000 |
| H | 2.68129800 | -2.27693500 | -0.11112600 |
| C | 2.11900100 | -5.51609700 | 0.86038800 |
| H | 3.91099600 | -4.33579800 | 0.47625700 |
| C | 0.76522900 | -5.53875200 | 0.89781000 |
| H | 2.68369200 | -6.40693100 | 1.10910000 |
| H | 0.22882700 | -6.43957300 | 1.17392800 |
| C | -1.37672700 | -4.35986400 | 0.60971900 |
| H | -1.93698000 | -5.24726600 | 0.88085700 |